\documentclass[aip,jcp,preprint]{revtex4-1}
\usepackage{graphicx}
\usepackage{rotating}
\usepackage{amsfonts,amsmath,latexsym,bbm,dsfont}

\usepackage{color}

\newcommand{\nl}{\nonumber \\ & }

\newcommand{\beq}{\begin{eqnarray}}
\newcommand{\eeq}{\end{eqnarray}}
\newcommand{\half}{\frac{1}{2}}

\newcommand{\op}{\hat}
\newcommand{\mat}{\mathbf}

\newcommand{\el}{}

\newcommand{\tab}[1]{Table~\ref{#1}}
\newcommand{\dg}{\dagger}
\newcommand{\lk}{\left}
\newcommand{\rk}{\right}

\newcommand{\Perm}[2]{\hat{\cal P}^{#1}_{#2}}

\def\RCS$#1: #2 ${\@namedef{RCS#1}{#2}\typeout{RCS #1: #2}}
\mathchardef\lt="313C \mathchardef\gt="313E
\mathcode`<="4268 \mathcode`>="5269

\renewcommand\vec\mathbf

\begin{document}

\author{Thomas Schraivogel}
\email{t.schraivogel@fkf.mpg.de}
\author{Aron J. Cohen}
\author{Ali Alavi}
\author{Daniel Kats}
\email{d.kats@fkf.mpg.de}
\affiliation{
Max Planck Institute for Solid State Research, Heisenbergstra\ss e 1, 70569 Stuttgart, Germany}

\title{Transcorrelated coupled cluster methods}
\begin{abstract}
Transcorrelated coupled cluster and distinguishable cluster methods are presented.
The Hamiltonian is similarity transformed with a Jastrow factor in the first quantisation,
which results in up to three-body integrals. The coupled cluster with singles and doubles equations
on this transformed Hamiltonian are formulated and implemented. 
It is demonstrated that the resulting methods have a superior basis set convergence and accuracy
to the corresponding conventional and explicitly correlated methods.
Additionally, approximations for three-body integrals are suggested and tested.
\end{abstract}

\maketitle 
\section{Introduction}

An accurate description of the dynamical electron correlation usually requires large one-particle basis sets 
and high-dimensional tensors to represent the correlated movement of electrons. 
Various approaches have been developed to accelerate the basis-set and excitation-level convergence. 
Explicit correlation methods introduce explicit dependence on the electron-electron distance 
into the wavefunction, which drastically reduces the finite basis-set error. 
\cite{k1985,kk1991i,ks2002,m2003,tenno2004,tenno2004a,v2004,tk2005,kedzuch:05,fkh05,fhk06b,f12g,noga:07,ccf12,
tknh07,rmp2f12,skhv08a,skhv08b,nkst08,tkh2008,valeev08,valeev08b,Torheyden:2008,bokhan2008,kaw2009,Werner:2011,
Ten-no:TCA131-1,Hattig:CR112-4,Kong:CR112-75}
The coupled cluster hierarchy \cite{cizek:66,Purvis:82,Raghavachari:89} is known to converge extremely fast 
to the full configuration interaction (FCI) results as long as the electron correlation is weak. 
\cite{tajti_heat:_2004,bomble_high-accuracy_2006,harding_high-accuracy_2008} 

However, both methodologies have limitations. Efficient implementations of the explicitly correlated (F12) methods
rely on various approximations, and the formalism is difficult to extend to higher than doubles excitation
classes. \cite{kohnExplicitly2010} The fixed-amplitude ansatz commonly employed in the F12 methods makes the methods less suitable
for pairs containing core orbitals, although the problem can be somewhat mitigated by using different length scales in 
the correlation factor for valence and core electrons \cite{Werner:2011} or by partially relaxing 
the F12 amplitudes \cite{tew_relaxing_2018}.
The coupled-cluster methods are sensitive to strong electron correlation and results are often not even 
qualitatively correct if large amount of static correlation is involved. 
Besides, at least perturbative triple excitations \cite{Raghavachari:89} are needed for accurate results, 
which makes the calculations expensive.  
Modified coupled cluster methods exist which demonstrate higher accuracy for a given excitation level.
\cite{Meyer:71,paldus_approximate_1984,piecuch_solution_1991,piecuch_approximate_1996,kowalski_method_2000,
bartlett_addition_2006,nooijen_orbital_2006, Neese:09, huntington_pccsd:_2010,robinson_approximate_2011,
huntington_accurate_2012,paldusExternally2017,black_statistical_2018,kats_dc_2013,kats_dcsd_2014,kats_accurate_2015,katsDistinguishable2019,rishiCan2019}
The distinguishable cluster approach is one of these methods, and is known to improve not only the accuracy for 
weakly correlated systems, but also to yield qualitatively good results for many strongly correlated systems.
\cite{kats_dc_2013,kats_dcsd_2014,kats_accurate_2015,kats_distinguishable_2016,kats_improving_2018}

An alternative route to improve the accuracy of electron-correlation treatment is to apply 
a similarity transformation to the Hamiltonian, which can be used to incorporate some correlation 
effects into the transformed Hamiltonian and thus simplify the problem for the electron-correlation methods. 
This idea goes back to the pioneering work of Boys and Handy and is termed \textit{transcorrelation}, \cite{boysCalculation1969}  
and it has been recently demonstrated that a combination of the transcorrelation with a stochastic 
full configuration interaction quantum Monte-Carlo (FCIQMC) \cite{Booth2009,Cleland2010,Booth2014} yields very promising results 
for weakly and strongly correlated systems.\cite{luoCombining2018,dobrautzCompact2019,cohenSimilarity2019,gutherberyliumdimer2021}
Since the Hamiltonian itself is transformed, the transcorrelation approach is not specific for 
low-order methods and can be used together with any excitation orders. 
The transcorrelated Hamiltonian has been applied before to linearized coupled cluster with singles and doubles, \cite{hinoApplication2002}
and for the uniform electron gas for coupled cluster and distinguishable cluster with singles and doubles, \cite{liaoEfficient2021}
however with some approximations to the three-body terms in the transformed Hamiltonian and with a simpler 
correlation factor.
In this communication we explore the quality of full transcorrelated coupled cluster and distinguishable 
cluster methods as well as their approximated versions.

\section{Theory}
\subsection{Transcorrelated coupled cluster}
\label{sec:tc-cc}

The transcorrelated formalism starts by similarity transforming the electronic Hamiltonian with the 
Jastrow factor,
\begin{align}
\exp(\tau)&=\exp{\sum_{\el i \lt \el j} u(\mat r_{\el i},\mat r_{\el j})}
\label{eq:jastrow}
\end{align}
 which can be efficiently done \textit{before} going to the second-quantisation formulation. 
$u(\mat r_{\el i},\mat r_{\el j})$ is a symmetric correlation function, 
$u(\mat r_{\el i},\mat r_{\el j})=u(\mat r_{\el j},\mat r_{\el i})$. 
The specific form of the correlation factor can be adjusted to the problem under consideration. 
We will be using the Boys-Handy correlation factor,\cite{boysCalculation1969} \textit{vide infra}.
Since the Coulomb operator commutes with $\tau$, and $\tau$ depends only on positions of two electrons,
the Backer-Campbell-Hausdorff expansion of the similarity transformation naturally truncates 
at the second commutator, and the original Hamilton operator is augmented with at most three-body 
terms. 
The transformed non-hermitian Hamiltonian in the second-quantised form becomes
\begin{align}
&\tilde H = \sum_{pq}h_p^q \sum_{\sigma} a^\dg_{p\sigma}a_{q \sigma} +
\half\sum_{pqrs}\lk(V_{pr}^{qs} - K_{pr}^{qs}\rk)\sum_{\sigma\rho}a^\dg_{p\sigma} a^\dg_{r\rho} a_{s \rho} a_{q \sigma}\nl
-\frac{1}{6}\sum_{pqrstu}L_{prt}^{qsu}\sum_{\sigma\rho\tau}a^\dg_{p\sigma} a^\dg_{r\rho}a^\dg_{t\tau} a_{u \tau} a_{s \rho} a_{q \sigma},
\label{eq:tch}
\end{align}
with one ($h_p^q = <p|\op h|q>$) and two- ($V_{pr}^{qs}=<pr|qs>$) electron part of the original electronic Hamiltonian,
and the additional terms arising from the similarity transformation,
\begin{align}
&K_{pr}^{qs}=\half<pr|\Perm{\el i}{\el j}\lk(\nabla^2_{\el i} u(\mat r_{\el i},\mat r_{\el j}) +(\nabla_{\el i}u(\mat r_{\el i},\mat r_{\el j}))^2
+2(\nabla_{\el i}u(\mat r_{\el i},\mat r_{\el j})\cdot \nabla_{\el i})\rk)|qs>,\nl
L_{prt}^{qsu}=<prt|\Perm{\el i}{\el j\el k}\lk(\nabla_{\el i} u(\mat r_{\el i},\mat r_{\el j})\cdot 
\nabla_{\el i} u(\mat r_{\el i},\mat r_{\el k})\rk)|qsu>.
\label{eq:tcterms}
\end{align}
$\Perm{\el i}{\el j}$ and $\Perm{\el i}{\el j \el k}$ are permutation operators, e.g.,
\begin{align}
\Perm{\el i}{\el j\el k}X_{\el i \el j \el k}=X_{\el i \el j \el k}+X_{\el j \el i \el k}+X_{\el k \el j \el i}
\end{align}
Note that the three-body operator is hermitian, and for real orbitals possesses 48-fold symmetry.

The transformed electronic Schr\"odinger equation,
\begin{align}
&\tilde H \Psi = E \Psi
\label{eq:schroed}
\end{align}
can approximately be solved using coupled-cluster methods. We restrict ourselves to singles and doubles.
However, we would like to stress that the transcorrelated Hamiltonian can be used together with 
coupled-cluster methods truncated at any excitation level. 
The coupled-cluster amplitude equations including three-body integrals can be derived using the well-known
second-quantisation algebra. The operators are normal-ordered which yields effective 0, 1, and 2-body
contributions from the three-body integrals. Singles-dressed integrals can be used to simplify the 
amplitude equations.
The three-body terms lead to increase of the computational scaling 
with respect to the system size from ${\cal O}(N^6)$ to ${\cal O}(N^7)$, but as with the conventional 
Hamiltonian, at the CCSD level at most quadratic doubles terms appear in the amplitude equations. 
%The transcorrelated CCSD (TC-CCSD) amplitude equations can be found in Appendix.
In the transcorrelated DCSD (TC-DCSD) method only terms involving effective two-body integrals
are modified according to the usual distinguishable cluster approximation.

We evaluate two types of approximations for the three-body integrals. 
In the first approximation (denoted in the following as approximation A), 
only terms involving explicit normal-ordered dressed three-body integrals 
are neglected, i.e., the  effective 0, 1, and 2-body terms are calculated using dressed three-body integrals
and contribute to the amplitude equations.
This approximation corresponds to neglecting three-body integrals normal-ordered  with respect to 
``optimized'' orbitals. 
In the second type of approximation (approximation B), only three-body contributions to the 0-2 body terms 
\textit{before dressing} are retained, i.e., this approximation corresponds to neglecting three-body terms 
normal-ordered with respect to the HF determinant.
This approach allows for very efficient implementation of transcorrelated methods, since the three-body terms 
can be self-contracted at the construction time on-the-fly and stored together with effective 0-2 body integrals. 

\section{Test calculations}
\label{sec:results}
The TC-CCSD and TC-DCSD amplitude equations have been implemented using Integrated Tensor Framework (ITF)
in Molpro.\cite{molpro} 
The integrals are calculated as outlined in Ref.~\onlinecite{cohenSimilarity2019} and imported using an FCIDUMP-type
interface.  

We employ the same Boys-Handy correlation factors as in Ref.~\onlinecite{cohenSimilarity2019},
\begin{align}
u(\mat r_{\el i}, \mat r_{\el j})&=\sum_{mno} c_{mno} \Perm{\el i}{\el j}\bar r_{\el i}^m \bar r_{\el j}^n
\bar r_{\el i \el j}^o, \quad \bar r=\frac{r}{1+r},
\label{eq:BH}
\end{align}
with 17 $c_{mno}$ parameters optimized at the variational Monte-Carlo (VMC) level taken from Ref.~\onlinecite{schmidtCorrelated1990}.

The calculations have been done using the cc-pVTZ basis set. The corresponding numbers for cc-pVDZ basis
set can be found in the supplementary material. 
The explicitly correlated results have been obtained using unrestricted CCSD-F12a, DCSD-F12a and 
CCSD(T)-F12a methods based on the restricted open-shell HF reference. \cite{kaw2009,kats_accurate_2015}
In all calculations all electrons have been correlated.
The results are compared to experimental numbers back-corrected for relativistic effects.\cite{chakravortyGroundstate1993,cohenSimilarity2019}

First, we compare the absolute atomic energies to TC-FCIQMC and experimental numbers, \tab{tab:energies}.
Evidently, transcorrelation drastically improves the basis set convergence. The results become very 
close to experimental values. The explicitly correlated coupled cluster numbers (using F12a approximations) 
are also given for comparison. It is obvious that the TC methods are much more accurate. 
The mean absolute errors of the transcorrelated methods compared to experimental numbers are 3-3.5 times 
smaller than of the explicitly correlated methods, and six times smaller than of the conventional methods 
(\tab{tab:ref_energies}).

Even more interesting is the ability of the transcorrelation to improve the accuracy of the underlying 
method itself.
For all atoms TC-CCSD differs from TC-FCIQMC by less than 1~m$E_h$. This can be compared to the accuracy 
of the conventional CCSD versus FCI, \tab{tab:ref_energies}, where CCSD shows errors of over 4~m$E_h$. 
The same boost in accuracy can be seen for DCSD. The maximal deviation of TC-DCSD from TC-FCIQMC 
is 0.1~m$E_h$, and the largest discrepancy of DCSD from FCI is more than 1~m$E_h$. 
Even CCSD(T) has a larger maximal error of 0.5~m$E_h$.
Thus, transcorrelation can simplify electronic structure problems, and low-level methods 
with only up to connected doubles excitations can reach the accuracy of more expensive higher-order methods. 

The approximate TC-CCSD and TC-DCSD methods with some of the terms involving the three-body integrals 
neglected are also very accurate. The results from approximation A are in nearly all cases closer to 
the corresponding complete transcorrelated methods with maximal deviations of 0.1~m$E_h$, whilst 
the approximation B causes deviations of up to 0.25~m$E_h$.
However, overall the gain from the transcorrelation is striking even with these approximations.

\begin{table*}[htbp]
\caption{Absolute energies of the neutral atoms in the ground-state in hartree using cc-pVTZ basis set. Approx. A denotes TC-CC without normal-ordered dressed three-body terms.
         Approx. B denotes TC-CC without normal-ordered three-body terms.} %TODO. %TODO SM17 is the correlation factor from}
\begin{ruledtabular}
\footnotesize
\begin{tabular}{lccccccccc}
Method      & Li       & Be        & B         & C         & N         & O         & F         & Ne         & MAE   \\
TC-CCSD     & -7.47803 & -14.66788 & -24.64922 & -37.83829 & -54.57901 & -75.05232 & -99.71314 & -128.90888 & 0.011 \\
Approx. A   & -7.47803 & -14.66795 & -24.64920 & -37.83826 & -54.57896 & -75.05223 & -99.71305 & -128.90884 & 0.011 \\
Approx. B   & -7.47801 & -14.66797 & -24.64925 & -37.83829 & -54.57895 & -75.05207 & -99.71297 & -128.90883 & 0.011 \\
TC-DCSD     & -7.47803 & -14.66787 & -24.65013 & -37.83937 & -54.57981 & -75.05292 & -99.71377 & -128.90947 & 0.010 \\
Approx. A   & -7.47803 & -14.66794 & -24.65011 & -37.83934 & -54.57976 & -75.05282 & -99.71367 & -128.90942 & 0.010 \\
Approx. B   & -7.47801 & -14.66796 & -24.65016 & -37.83937 & -54.57975 & -75.05267 & -99.71360 & -128.90942 & 0.010 \\
CCSD-F12    & -7.47258 & -14.65752 & -24.63682 & -37.81916 & -54.55336 & -75.01842 & -99.67121 & -128.86065 & 0.035 \\
DCSD-F12    & -7.47258 & -14.65760 & -24.63825 & -37.82108 & -54.55521 & -75.02070 & -99.67400 & -128.86395 & 0.034 \\
CCSD(T)-F12 & -7.47259 & -14.65774 & -24.63839 & -37.82134 & -54.55565 & -75.02134 & -99.67475 & -128.86477 & 0.033 \\
TC-FCIQMC   & -7.47804 & -14.66789 & -24.65003 & -37.83928 & -54.57989 & -75.05303 & -99.71377 & -128.90944 & 0.010 \\
Expt.       & -7.47806 & -14.66736 & -24.65391 & -37.84500 & -54.58920 & -75.06730 & -99.73390 & -128.93760 &     
\end{tabular}
\label{tab:energies}
\end{ruledtabular}
\end{table*}

\begin{table*}[htbp]
\caption{Absolute energies of the neutral atoms in the ground-state in hartree using cc-pVTZ basis set.}
\begin{ruledtabular}
\begin{tabular}{lccccccccc}
Method   & Li       & Be        & B         & C         & N         & O         & F         & Ne         & MAE     \\
CCSD    & -7.44605 & -14.62356 & -24.60376 & -37.78725 & -54.52246 & -74.98187 & -99.62848 & -128.81081 & 0.063 \\
DCSD    & -7.44605 & -14.62366 & -24.60526 & -37.78927 & -54.52441 & -74.98425 & -99.63136 & -128.81419 & 0.061 \\
CCSD(T) & -7.44607 & -14.62379 & -24.60538 & -37.78953 & -54.52487 & -74.98494 & -99.63220 & -128.81513 & 0.061 \\
FCI      & -7.44607 & -14.62381 & -24.60582 & -37.79004 & -54.52524 & -74.98528 & -99.63243 & -128.81521 & 0.061 \\
Expt.    & -7.47806 & -14.66736 & -24.65391 & -37.84500 & -54.58920 & -75.06730 & -99.73390 & -128.93760 &       
\end{tabular}
\label{tab:ref_energies}
\end{ruledtabular}
\end{table*}

The accuracy of relative energies has been evaluated by computing atomic ionisation potentials (IPs), \tab{tab:ip}.
The IPs from transcorrelated methods are very close to the experimental values 
with the mean absolute errors of only 0.64, 0.56, and 0.58 m$E_h$ 
and maximal deviations of 2.3, 2.1, and 1.5 m$E_h$ (in the O case) for TC-CCSD, TC-DCSD and TC-FCIQMC, respectively.
Errors of the explicitly correlated coupled cluster methods are much larger, 
with the mean absolute errors of 3.7, 3.0, and 2.9 m$E_h$ and  
with maximal deviations of 8.2, 6.8, and 6.6 m$E_h$ (in the Ne case) for CCSD-F12a, DCSD-F12a and CCSD(T)-F12a,
respectively.
The errors of conventional methods using cc-pVTZ basis are even larger with mean absolute 
errors ranging from 5.6 m$E_h$ for FCI to 6.6 m$E_h$ for CCSD.  
Note that in the case of transcorrelated methods the HF orbitals and Jastrow factors are optimized for the neutral atoms, 
and the same integrals are reused for the cations, i.e.,
neither orbitals nor Jastrow factors are reoptimized for the cations. Thus, some bias 
towards the neutral atoms is expected in our results.
However, the good agreement with the experimental values and TC-FCIQMC suggests that the partial orbital 
relaxation coming from the single excitations is sufficient to largely eliminate this problem, 
and demonstrates the transferability of the approach, even for methods truncated at the singles and doubles level.

As in the case of absolute energies, the IPs from transcorrelated CCSD 
are closer to the corresponding FCI reference values, \tab{tab:ip}, than for the conventional CCSD method,
\tab{tab:ref_ip}, although the accuracy gain is smaller here. The largest deviations of TC-CCSD from TC-FCIQMC IPs is 0.87 m$E_h$,
and of CCSD from FCI IPs is 1.77 m$E_h$.  

The results from the approximate transcorrelated methods are also encouraging. 
The approximation A changes the transcorrelated results by at most 0.5 m$E_h$ (for the Ne atom). 
The maximal deviation of the approximation B from the complete transcorrelated methods is 1.7 m$E_h$,
which is largely due to the nonoptimal HF reference for the cation used for the normal-ordering, \textit{vide supra}.
However, TC-CCSD with only self-contracted three-body terms 
(Approximation B) benefits in this case from the error cancellation and is closest to experimental IPs, 
with errors smaller than from TC-FCIQMC for nearly all first-row atoms, and the mean absolute error of 
only 0.41 m$E_h$.
The errors caused by the three-body approximations are very similar for TC-CCSD and TC-DCSD, which 
hints to a possibility to treat the full three-body terms at a low level of theory and 
use it as a correction for high-level coupled cluster methods.
\begin{table*}[htbp]
\caption{IPs in m$E_h$ using cc-pVTZ basis set. Approx. A denotes TC-CC without normal-ordered dressed three-body terms.
         Approx. B denotes TC-CC without normal-ordered three-body terms.} %TODO. %TODO SM17 is the correlation factor from}
\begin{ruledtabular}
\begin{tabular}{lccccccccc}
Method      & Li     & Be     & B      & C      & N      & O      & F      & Ne     & MAE   \\
TC-CCSD     & 198.54 & 342.48 & 304.83 & 413.81 & 534.73 & 498.18 & 640.43 & 793.32 & 0.64 \\
Approx. A   & 198.54 & 342.55 & 304.85 & 413.84 & 534.67 & 497.95 & 640.03 & 792.79 & 0.76 \\
Approx. B   & 198.50 & 342.32 & 304.94 & 414.32 & 535.29 & 499.12 & 641.06 & 794.64 & 0.41 \\
TC-DCSD     & 198.54 & 342.47 & 305.74 & 414.10 & 534.67 & 498.44 & 640.81 & 793.83 & 0.56 \\
Approx. A   & 198.54 & 342.54 & 305.76 & 414.12 & 534.61 & 498.21 & 640.41 & 793.30 & 0.69 \\
Approx. B   & 198.50 & 342.31 & 305.84 & 414.60 & 535.24 & 499.37 & 641.43 & 795.13 & 0.61 \\
CCSD-F12    & 197.75 & 342.74 & 302.63 & 411.55 & 531.58 & 494.47 & 633.96 & 786.27 & 3.72 \\
DCSD-F12    & 197.75 & 342.81 & 303.98 & 412.20 & 531.90 & 495.46 & 635.20 & 787.70 & 2.98 \\
CCSD(T)-F12 & 197.76 & 342.94 & 303.95 & 412.33 & 532.12 & 495.76 & 635.43 & 787.91 & 2.86 \\
TC-FCIQMC   & 198.54 & 342.64 & 305.66 & 414.66 & 535.31 & 499.05 & 640.95 & 793.95 & 0.58 \\
Expt.       & 198.15 & 342.58 & 304.99 & 413.97 & 534.60 & 500.50 & 641.10 & 794.50 &      
\end{tabular}
\label{tab:ip}
\end{ruledtabular}
\end{table*}
\begin{table*}[htbp]
\caption{IPs in m$E_h$ using cc-pVTZ basis set.}
\begin{ruledtabular}
\begin{tabular}{lccccccccc}
Method   & Li     & Be     & B      & C      & N      & O      & F      & Ne     & MAE  \\
CCSD    & 196.70 & 340.98 & 301.01 & 410.18 & 530.41 & 488.36 & 628.42 & 781.39 & 6.62 \\
DCSD    & 196.70 & 341.08 & 302.41 & 410.88 & 530.75 & 489.39 & 629.69 & 782.85 & 5.83 \\
CCSD(T) & 196.71 & 341.20 & 302.37 & 411.02 & 531.00 & 489.71 & 629.96 & 783.10 & 5.66 \\
FCI      & 196.71 & 341.22 & 302.78 & 411.21 & 531.01 & 489.82 & 629.99 & 783.04 & 5.58 \\
Expt.    & 198.15 & 342.58 & 304.99 & 413.97 & 534.60 & 500.50 & 641.10 & 794.50 &     
\end{tabular}
\label{tab:ref_ip}
\end{ruledtabular}
\end{table*}

\section{Conclusions}
The transcorrelated approach combined with the coupled cluster methods shows great promise. Not only
does it drastically improve the basis set convergence, but it can also increase the accuracy of the CC method 
itself. 
The non-hermiticity of the transcorrelated Hamiltonian does not cause problems. 
The high degree of orbital invariance of coupled cluster methods allows to use conventional HF 
orbitals instead of solving bi-orthogonal HF equations.

Our results demonstrate that the expensive three-body terms can be approximated by neglecting their normal-ordered contributions.
The self-contraction of the integrals can be done on-the-fly, and the resulting effective 0 - 2 body
contributions can be stored together with other terms of the Hamiltonian. The quality of the approximation
can be improved by using better orbitals, e.g., Brueckner orbitals, as shown by the singles-dressed
formulation (approximation A). 

The transcorrelated approach can be easily applied to higher order coupled cluster methods,
especially if approximated three-body terms are used. Moreover, the use of highly flexible Jastrow functions, which incorporate information on the position of the nuclei as well as electron-electron distances, can be expected to bring further benefits in the treatment of more complex systems. 

\begin{acknowledgements}
Funded by the Deutsche Forschungsgemeinschaft (DFG, German Research Foundation) -- 455145945. Financial support from the Max-Planck Society is gratefully acknowledged. \end{acknowledgements}

\bibliography{methods,molpro,dc,tensor,f12,fciqmc,tc}

\end{document}